\documentclass{optica-article}

\journal{opticajournal} 

\articletype{Research Article}

\usepackage{placeins}
\usepackage{textgreek}

\begin{document}

\title{In-Situ Measurement of Beam Divergence in a High Efficiency SNSPD Platform}

\author{Daniel W. Sorensen,\authormark{1} \authormark{2}, Martin J. Stevens,\authormark{2}
 Lynden K. Shalm~\authormark{1}~\authormark{2}~\authormark{3}, and  Dileep V. Reddy\authormark{2} \authormark{3}}

\address{\authormark{1}Department of Electrical, Computer and Energy Engineering, University of Colorado, Boulder, CO USA\\
\authormark{2}Physical Measurement Laboratory, National Institute of Standards and Technology, Boulder, CO USA\\
\authormark{3}Department of Physics, University of Colorado, Boulder, CO USA\\}

\begin{abstract*} 
We implement a time-of-flight imaging technique utilizing a differential-readout SNSPD to spatially resolve detection events in a fiber-coupled detector platform. We measure the spatial detection profiles for ultra-high numerical aperture fiber, standard single-mode fiber, and thermally-expanded core fiber (mode-field diameters 4.1$\mu$m, 10.4$\mu$m, 30$\mu$m respectively) in an active area surrounded by an all-dielectric optical stack designed for near-unity detection efficiency. We see no beam divergence in all but the smallest fiber optic modes. This contradicts previously-held beliefs that beam divergence during the detection process necessitates active areas much larger than coupled optical modes, opening new paths toward smaller and better-optimized detectors.
\end{abstract*}

\section{Introduction}
The near-unity system detection efficiency (SDE) offered by modern superconducting nanowire single-photons detectors (SNSPDs) enables a range of quantum optics applications that rely on the extremely efficient counting of photons \cite{shalm_2015,giustina_2015,zhong_2020,liu_2023}. In fiber-coupled systems this can be accomplished by embedding the superconducting active area in a highly optimized dielectric stack that is designed to interferometrically trap incident photons to maximize detection probability. While this is a proven method to achieve near-unity SDE \cite{marsili_2013, reddy_2020}, there are outstanding questions regarding the impact that it has on the spatial profile of detected light. It has been suggested that, as these dielectric structures cause incident photons to pass through the thin superconducting active area many times, the accumulated path length is enough to cause significant beam divergence of incident fiber optic modes \cite{reddy_2020}. This would require SNSPD active areas to be much larger than their incident fiber optic modes in order to collect all incident light -- a sacrifice that hurts the timing jitter, maximum count rate, and fabrication yield of such devices \cite{lita_2022,jaha_2025}. Here we implement an electrical time-of-flight (TOF) imaging technique to investigate the spatial distribution of detection events in such a platform. Using an SNSPD designed to image the spatial distribution of detection events along one axis, we can measure the degree to which this hypothesized beam divergence plays a role in the detection process.

Single-photon sensitivity in superconducting microbridges -- ultra-thin layers of superconducting material patterned into microscale strips -- was first demonstrated in 2001 \cite{Goltsman_2001}. The technology immediately showed promise for applications that demand single-photon detection with high timing resolution, low timing jitter, and low dark count rates. While early devices were limited to very low detection efficiencies in the near-infrared \cite{hadfield_2005, verevkin_2002}, evidence suggested that this was not an intrinsic limitation of the platform. Demonstrations of devices with high internal quantum efficiency \cite{anant_2008} and devices with a saturated internal quantum efficiency over a wide range of wavelengths and bias currents \cite{baek_2011} indicated that advances in fabrication techniques and material science could yield large meandered active areas with high, uniform photosensitivity. It was thus only an engineering challenge to overcome the obstacles standing in the way of high detection efficiency -- primary among them being the low absorption probability that a photon experiences during a single pass through a detector's superconducting layer, which is only nanometers thick. 

Following a rapid succession of record-breaking SDE demonstrations enabled by careful optical engineering \cite{verevkin_2004,rosfjord_2006,verma_2012}, the state-of-the-art reached 93\% SDE in 2013 \cite{marsili_2013}.  This approach overcame the low single-pass optical absorption of the thin superconducting films used in SNSPDs by embedding them in an optical structure composed of highly reflective layers below and anti-reflective layers atop the active area optimized to maximize normal-incidence absorption at the target wavelength. These structures trap incident light, causing photons to repeatedly pass through the superconducting layer until absorption probability approaches unity.  One demonstration that surpassed this record \cite{reddy_2020} replaced the metallic reflective layer in reference \cite{marsili_2013} with a 13-layer dielectric distributed Bragg reflector (DBR), eliminating nearly 3\% of loss attributed to non-zero resistance of the reflecting conductor. This resulted in a greatly expanded dielectric stack nearly 3~\textmu m thick and a new record SDE of 98\% at a wavelength of 1550 nm.  This breakthrough was coupled with results indicating a significant gain in SDE as the active area diameter is increased from 20~\textmu m to 35 ~\textmu m when illuminated by SMF-28e+ fiber with a mode-field diameter (MFD) around 10~\textmu m. Gaussian beam expansion was proposed as an explanation for this discrepancy. The relatively large optical path length accumulated during multiple round trips in this thick optical stack was thought to be comparable with the $\sim$50~\textmu m Rayleigh range of this incident mode, causing coupled optical modes to expand greatly during the detection process.

Understanding the spatial profile of detection events in a given high SDE platform is essential for the optimization of next-generation SNSPDs. A well-optimized active area must be large enough to reliably capture a near-unity proportion of this spatial profile without sacrificing performance to meandered regions that receive no incident light. While high-efficiency single- and multi-pixel SNSPD structures with large active areas are common in the literature \cite{wollman_2024, zhang_2019, lv_2017}, every unit length of nanowire increases the kinetic inductance of the device. This degrades device performance in terms of maximum count rate and timing jitter \cite{lita_2022}, two metrics that are essential in many present and future SNSPD application settings. To push performance on these fronts while maintaining near-unity SDE we thus need a method to better understand the spatial profile of detection events and the impact that different optical stack platforms can have on it.

Our method of spatially resolving detection events using a specialized SNSPD addresses this need. By performing \textit{in situ} imaging of detection events we can investigate the degree to which hypothesized beam divergence contributes to the detection process in an all-dielectric platform designed for high-SDE detection. This yields valuable insights into the detection process and promises to be a useful tool in further investigations toward future, near-ideal SNSPD systems.

\section{Imaging Devices and Measurement Setup}
\begin{figure}[!htpb]
\centering
\includegraphics[width=0.95\textwidth]{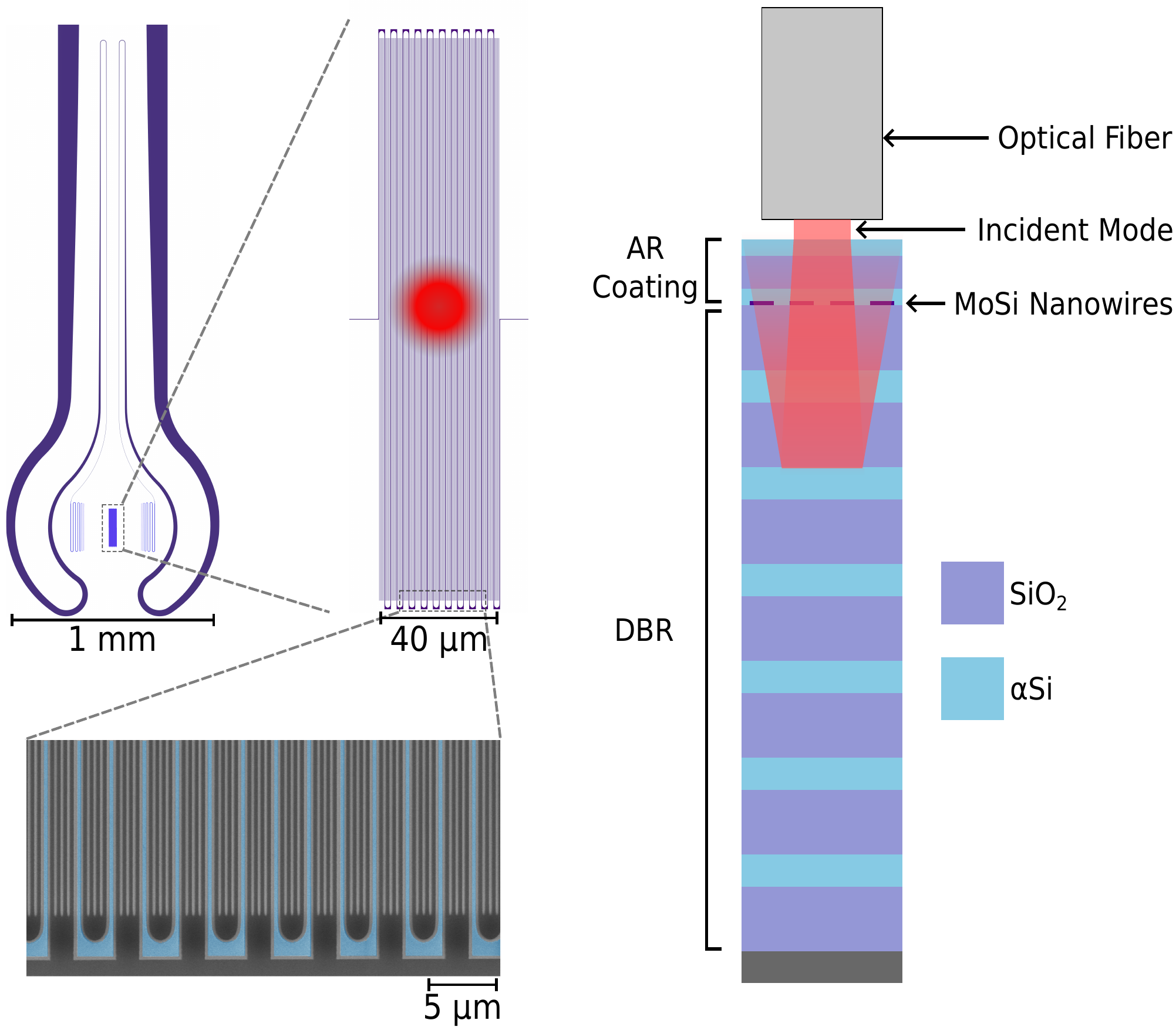}
\caption{\label{fig:device}Imaging device used for detection profile measurement. \textbf{(a):} Top-down view at different scales. Top left is a render of 10 mm coplanar waveguide Hecken taper that transforms the impedance of the nanowire to 50 $\Omega$ impedance of readout electronics. Grounding plane not shown for clarity. Top right is the imaging device active area, with approximate beam profile shown in red. 80~\textmu m of extra nanowire transmission lines on top and bottom delay pulses, allowing detection events on neighboring columns to be resolved temporally. Bottom is a false-color scanning-electron microscope image of a portion of the active area. 120 nm-wide current-carrying lines (highlighted in blue) are spaced 2.08~\textmu m apart to sample beam; ridges of MoSi etched in between mimic the optical properties of a 75\% fill-factor device while serving as a grounding plane for the current-carrying lines. \textbf{(b):} Side illustration of the 3~\textmu m thick dielectric stack surrounding the MoSi active area. 13 alternating $\lambda$/4 layers of SiO$_2$ and $\alpha$Si compose the DBR and 3 layers of the same materials compose the AR coating. Incident fiber-optic mode with exaggerated beam expansion shown in red.}
\end{figure}

\begin{figure}[!htpb]
\centering
\includegraphics[width=0.7\textwidth]{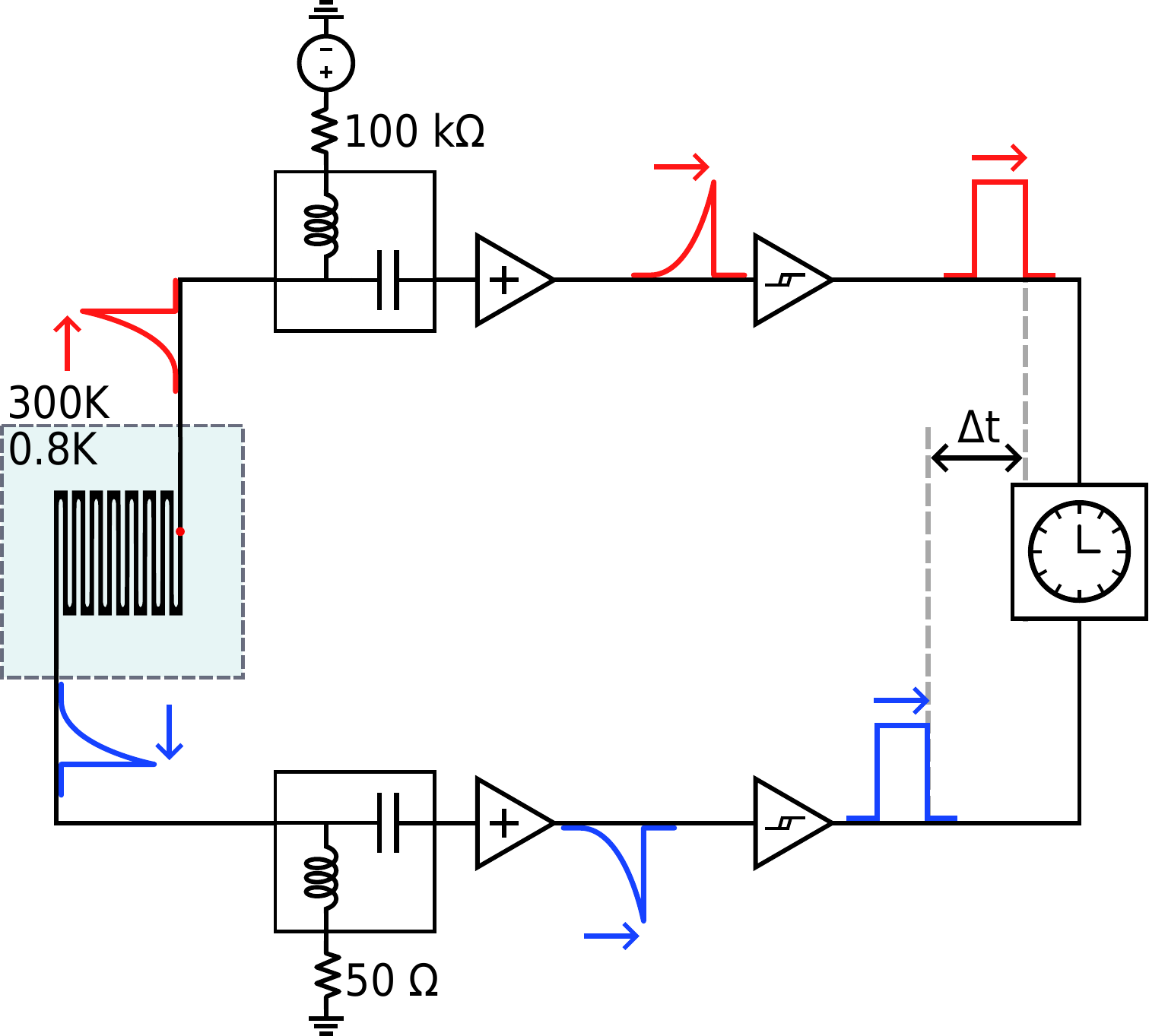}
\caption{\label{fig:circuit}Imaging device readout circuit schematic. Both ends of the differential readout device are connected to bias tees -- a quasicurrent is established by biasing the DC port of the `positive' end through a 100 k$\Omega$ series bias resistor and grounding the `negative' end through a 50 $\Omega$ termination load. AC signals are sent through amplifier and hysteretic discriminator circuits to translate them from mV-scale pulses of opposite polarity to binary signals suitable for timetagging electronics.}
\end{figure}

Detection events are imaged with differential-readout SNSPD devices (Fig. \ref{fig:device}) designed for TOF measurements that spatially resolve detection events along one axis of the active area. Each detection event generates two broadband, counter-propagating RF pulses of opposite polarity that travel along the nanowire. While measurement of only one of these pulses is enough to herald a detection event, measuring the timing difference between the two pulses as they arrive at their respective readout terminals allows us to infer the position along the length of the nanowire at which a detection event occurred \cite{hofherr_2013,calandri_2016,zhao_2017}. Current-carrying wires are spaced 2.08~\textmu m apart to sample the beam profile, and etched ridges in between with a period of 220 nm mimic the optical properties of a high fill-factor device. Devices are read out with 10 mm impedance-matching Hecken tapers at each end to adiabatically transform the $\sim$3k$\Omega$ nanowire impedance to the 50$\Omega$ readout impedance, helping to preserve the precise timing information in the pulses relevant to this measurement. Because the all-dielectric stack design allows for no grounding plane outside of the superconducting layer, all nanowire and taper transmission lines are in the coplanar waveguide configuration, with unetched superconducting material serving as the grounding plane with a gap width of 600 nm. Both device readouts are connected through a bias circuit (Fig. \ref{fig:circuit}) to time tagging electronics. 

Devices presented here follow a fabrication process similar to that in reference \cite{reddy_2020}; beginning with a 76.2 mm diameter silicon wafer, a 13 layer distributed Bragg reflector (DBR) is deposited using plasma-enhanced chemical vapor deposition (PECVD). Dielectric layers have optical thicknesses of $\lambda$ / 4 each, alternating between Silicon Dioxide (SiO$_2$, $n$ = 1.453) and amorphous Silicon ($\alpha$Si, n = 2.735). Optimzed for 1550 nm light, the SiO$_2$ and $\alpha$Si layers had nominal thicknesses of 266.7 nm and 141.7 nm respectively. Atop the DBR a 5 nm (Ti) / 50 nm (Au) / 5 nm (Ti) metallic layer is evaporatively deposited and patterned with a photolithographic liftoff process to create electrical bonding pads for differential device readout as well as various alignment marks for later lithographic layers. A $\sim$4.1 nm-thick superconducting layer of 75:25 ratio molybdenum silicide (MoSi, $T_c > 5K$ \cite{verma_2015}) layer is then deposited via magnetron sputtering and capped with a 2 nm layer of $\alpha$Si sputtered in-situ to prevent oxidization of the superconducting layer. Coarse photolithographic patterning for the impedance-matching tapers is etched into the MoSi layer via an SF$_6$ reactive-ion etch (RIE). Device active areas are patterned via electron-beam lithography and a PMMA resist process followed by the same RIE. Following this, a 3-layer 78.5 nm ($\alpha$Si) - 122.4 nm (SiO$_2$) - 66.7 nm ($\alpha$Si) anti-reflective (AR) coating is deposited on top through the same PECVD process. The nominal thicknesses of each layer are optimized to maximize absorption of $\lambda = 1550$ nm light for a given nanowire fill-factor. All lithographic patterns are generated with Phidl \cite{mccaughan_2021}.

The imaging devices are mounted in a standard fiber self-aligning mounting package \cite{miller_2011} and cooled in a sorption-based cryostat to 750-800 mK. These devices were coupled to three different optical fibers with differing MFDs to measure their respective detection profiles. The devices were illuminated with continuous-wave 1550 nm laser light attenuated to yield approximately $10^6$ counts per second on the imaging device. Compared to this count rate background counts were negligible.

\section{Results}
\begin{figure}[!htpb]
\centering
\includegraphics[width=\textwidth]{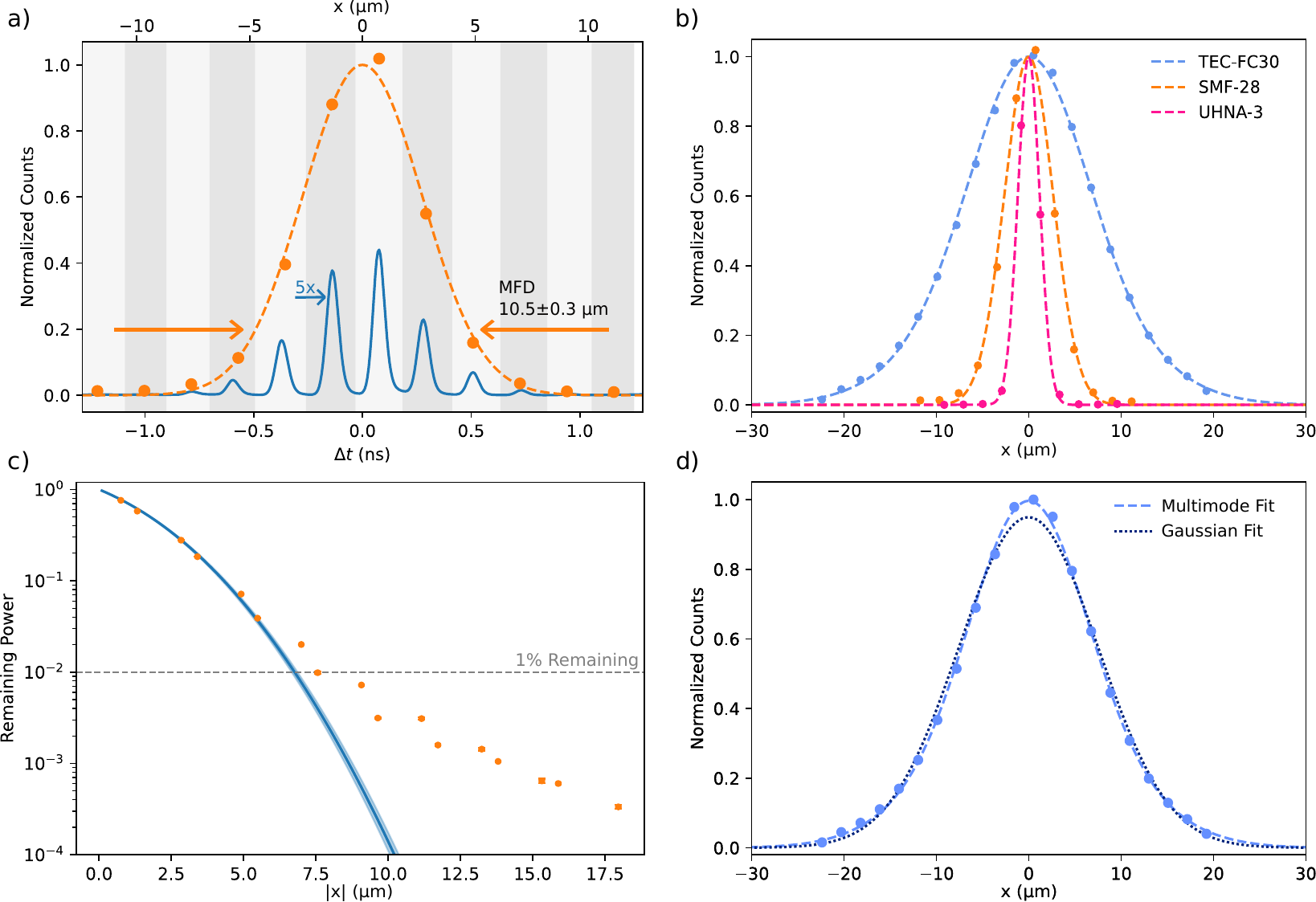}
\caption{\label{fig:timingtospatial}\textbf{(a):} Timing correlation between pulse pairs (blue) and binned spatial profile (orange) from imaging device illuminated by SMF-28e+ fiber (Manufacturer specified MFD = $10.4 \pm 0.5$~\textmu m, fit MFD = $10.5 \pm 0.3$~\textmu m). Timing curve exaggerated 5x for clarity. The timing bins corresponding to detection events in each device column are shaded. Each timing peak represents detection events along one column of the device. 215 ps average peak separation indicates an average pulse propagation velocity of $0.003$c. \textbf{(b):} Inferred spatial profile alongside fit of three different fiber optics: UHNA-3 (manufacturer specified MFD = $4.1 \pm 0.3$~\textmu m, fit MFD = $4.7 \pm 0.2$ ~\textmu m), SMF-28 (manufacturer specified MDF = $10.4 \pm 0.5$~\textmu m, fit MFD = $10.5 \pm 0.4$~\textmu m) and TEC-FC30 (manufacturer specified MFD = $30 \pm 2$~\textmu m, fit MFD = $29.4 \pm 0.2$~\textmu m). Gaussian modes suitably fit the spatial profiles of SMF-28 and UHNA-3, while the the $l = 0$, $p = 1$ Laguerre-Gaussian mode must be included to obtain a suitable fit for TEC-FC30. \textbf{(c):} Fractional power measured outside horizontal distance $|x|$ from center of SMF-28 beam. While measured data points (orange) agree with the fit Gaussian mode (dashed blue) near the center of the beam, about 1\% of power in the wings shows a wider spatial profile. \textbf{(d):} Comparison between Gaussian and multimode Laguerre-Gauss fits of TEC-FC30 spatial profile. While the multimode fit has most power in the fundamental $l = p = 0$ mode, the ideal fit indicates about 7\% of power is in the $l = p = 1$ mode.}

\end{figure}

When the devices are illuminated with fiber-coupled light we are able to clearly distinguish between detection events in neighboring columns through the timing correlation between pulses at each readout terminal. This allows us to bin counts by column and infer a marginal distribution of detection events along one axis of the device, sampled at the position of each current-carrying column (Fig. \ref{fig:timingtospatial}.a). Three different single-mode fibers were imaged: ultra-high numerical-aperture fiber (UHNA-3), standard telecom fiber (SMF-28), and thermally-expanded core fiber (TECFC-30) with manufacturer-specified MFDs of $4.1 \pm 0.3$~\textmu m, $10.4 \pm 0.5$~\textmu m, and $30 \pm 2$~\textmu m respectively (Fig \ref{fig:timingtospatial}.b). We fit the observed profiles in the basis of Laguerre-Gaussian modes: the observed profiles of UHNA-3 and SMF-28 are well fit with the lowest-order $l = p = 0$ mode (the fundamental Gaussian) except for a $\sim1\%$ discrepancy in the tails of the power spread (Fig \ref{fig:timingtospatial}.c). This could represent a small fraction of light that required multiple passes through the active area before absorption, or light scattered by the active area grating pattern. Regardless, it is not enough to explain the SDE gap between 20 and 35~\textmu m-diameter active areas reported in reference\cite{reddy_2020}. In contrast to the UHNA-3 and SMF-28 fibers, the TEC fiber has a significantly larger core diameter that results from an adiabatic expansion of a single-mode fiber at the end interfaces. The expanded end facets can in principle support higher-order modes. We find that the best fit to the measured profile contains a 93\% contribution from the fundamental Gaussian $l = p = 0$ mode and 7\% in the $l = 0$,  $p = 1$ mode, suggesting that light is coupled into the higher-order mode through imperfect core expansion (Fig \ref{fig:timingtospatial}.d).

The measured MFDs for UHNA-3, SMF-28 and TEC-FC30 were $4.7 \pm 0.2$~\textmu m, $10.5 \pm 0.4$~\textmu m, and $29.4 \pm 0.2$~\textmu m respectively. Of the three fiber optic modes imaged only the smallest, UHNA-3, shows measurable beam divergence. Neither of the larger fibers are measured to expand beyond their manufacturer-specified mode sizes. This is consistent with the scaling of the Rayleigh range with the square of the MFD, causing smaller modes to divergence significantly faster. Based on this measured divergence, we can infer the free-space equivalent optical path length that UHNA-3 light undergoes from fiber tip to detection to be less than 3~\textmu m. For all three fibers there is no evidence of beam divergence significant enough to cause an appreciable drop in SDE between a 20 and 35~\textmu m-diameter active area detector as speculated in reference \cite{reddy_2020}.

\section{Discussion and Outlook}
This data, for the first time, gives us an empirical look at the spatial profile of detection events in an all-dielectric stack SNSPD platform developed for near-unity SDE. Active areas in excess of 30~\textmu m in diameter have been thought necessary to achieve near-unity efficiency \cite{reddy_2020,reddy_2022}, and the beam divergence of SMF-28-coupled light was deemed a likely culprit. No such divergence is seen here. Moreover no measurable beam divergence is seen at all in SMF-28 light. A fit MFD = $10.5 \pm 0.3$~\textmu m indicates that a 20~\textmu m-diameter active area is sufficient for $>99.9\%$ coupling efficiency given a perfectly-aligned beam. We thus find no evidence that the divergence of incident light necessitates active areas much larger than their coupled optical modes. Further research is thus needed to explore mechanisms behind the results in \cite{reddy_2020} indicating the need for active areas $>30$~\textmu m in diameter.

\begin{figure}[!htpb]
\centering
\includegraphics[width=0.8\textwidth]{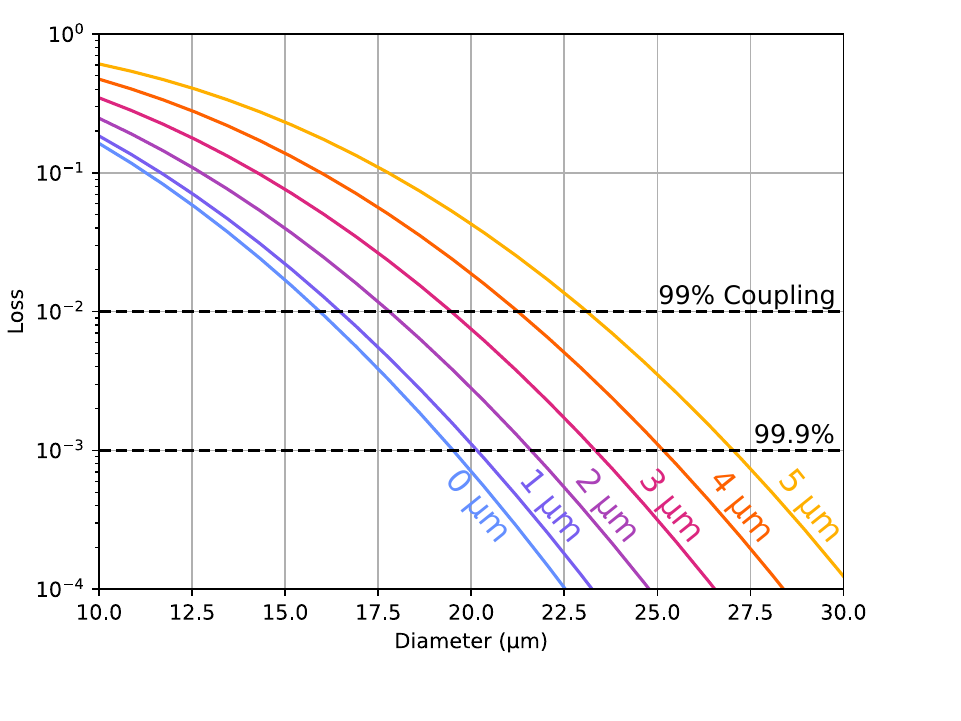}
\caption{\label{fig:e_vs_d} Simulated coupling loss vs. active area diameter for a gaussian SMF-28 input mode with MFD = 10.5~\textmu m. Each curve represents packaging misalignment values from 0 to 5~\textmu m. While a perfectly-aligned active area can achieve 99.9\% coupling efficiency with only a 20~\textmu m diameter, micron-scale misalignment quickly increases the required diameter to achieve the same coupling.}
\end{figure}

A possible explanation for these results is the increased sensitivity to fiber-detector misalignment that plagues smaller active areas. Micron-scale alignment errors introduced during the device mounting and fiber-coupling processes could introduce coupling losses that hurt the performance of an active area that would otherwise efficiently collect an incident beam. Active areas that extend well beyond an incident beam would be insensitive to such errors, while smaller active areas with no such overhead would see significant performance degradation. We can quantify this by computing the overlap of an offset gaussian mode with a circular active area of a given diameter. Doing so indicates that a 20$\mu m$ diameter active area can tolerate misalignment up to 3.4~\textmu m before coupling losses exceed 1\% for the measured SMF-28 beam profile; a 35~\textmu m active area can tolerate up to 11.2~\textmu m of misalignment. Assuming the magnitude of alignment errors seen in reference \cite{miller_2011}, this effect alone could explain the several-percent performance gap seen in reference \cite{reddy_2020}; at 4.5~\textmu m misalignment a 20~\textmu m active area would miss nearly 3\% of an incident SMF-28 mode while a 35~\textmu m active area still collects more than 99.9\% of incident light (Figure \ref{fig:e_vs_d}). The imaging technique used here offers a path to study this effect in more depth. Careful characterization of signal propagation times through each arm of readout electronics could allow for the absolute position of detection events in the active area to be inferred, providing direct measure of fiber-detector alignment. Statistical analysis of such fiber alignment measurements in many different devices could provide valuable insight into this effect and inform the development of techniques to reduce such alignment errors.

Our results show that excessive beam divergence is not a limiting factor the minimum diameter of efficient detectors. Consequently, it may be possible to develop smaller active area detectors that support both high count rates and near-unit detection efficiency. One approach would be to further reduce fiber-detector alignment errors to negate the effect discussed in the previous paragraph. More careful optimization of the micromachining method used in reference \cite{miller_2011} may yield tighter tolerances than are currently achieved, or a substrate different than the conventional silicon wafers currently used could be more resilient to errors introduced during the mounting process. Additionally, the minimal impact of beam divergence makes smaller fiber optic modes even more attractive. Even with the modest amount of divergence we measured, the 4.7~\textmu m MFD detection profile seen in light coupled through UHNA-3 fiber would dramatically reduce the requisite detector diameter compared to currently-used SMF-28; >99\% of a perfectly-aligned beam could be captured by a 7.5~\textmu m diameter active area, while the same coupling could be achieved by a 12.5~\textmu m active area subject to 4.5~\textmu m misalignment. While low-loss splicing between standard SMF and UHNA fibers remains challenging, progress is being made to approach conventional telecom splice losses \cite{yin_2019}.

This imaging technique could also prove helpful in exploring potential beam divergence in other SNSPD materials platforms. For example, demonstrations of near-infrared single-photon sensitivity in tungsten-silicide wires up to several microns wide \cite{chiles_2020,reddy_2022} rely on superconducting layers as thin as 2 nm, with a per-pass absorption probability likely much lower than that of our comparatively thick 4.1 nm MoSi superconducting layer. Dielectric structures optimized for efficient absorption in such microwire platforms may thus cause the effective fiber-to-detection optical path length to be substantially greater than that observed here, as more average passes through an active area are required for absorption probability to approach unity. As a result efforts toward high-efficiency photon counting in these microwire platforms would likely benefit from direct beam-profile characterization as presented here to explore potentially exaggerated beam divergence in these platforms.

\section{Conclusions}
We implemented a TOF imaging technique to characterize the spatial profile of detection events in an SNSPD platform featuring a thick dielectric stack designed for near-unity detection efficiency. While beam divergence is seen in a smaller fiber optic mode, no substantial divergence is seen in SMF-28 coupled light. This contradicts previous theories that the divergence of such light during the detection process necessitates active areas much larger than the incident optical mode, shifting the focus toward other factors such as fiber-detector misalignment in the pursuit of better-optimized SNSPDs capable of near-unity SDE.

\section{Acknowledgments}
The authors acknowledge Michael Grayson for fruitful lab discussions regarding this work, as well as Eli Mueller and Eric Rappeport for manuscript review.

\section{Funding}
The authors acknowledge support from the National Science Foundation QLCI OMA-2016244, the University of Colorado Quantum Engineering Initiative, and the National Institute of Standards and Technology. 


\FloatBarrier
\bibliography{BeamDivergence}

\end{document}